\documentclass[twocolumn,aps,prb,showpacs,amsmath,amssymb,reprint,superscriptaddress]{revtex4-1}
\usepackage[T1]{fontenc}
\usepackage[latin9]{inputenc}
\usepackage{amsmath}
\usepackage{graphicx}
\usepackage{color}

\begin{document}
\title{Anisotropic magnetic
excitations of a frustrated bilinear-biquadratic spin model
-- Implications for
spin waves of
 detwinned iron pnictides}
\author{Changle Liu}
\affiliation{State Key Laboratory of Surface Physics and Department of Physics, Fudan University, Shanghai, 200433, China}
\author{Xingye Lu}
\affiliation{Center for Advanced Quantum Studies and Department of Physics, Beijing
Normal University, Beijing 100875, China}
\author{Pengcheng Dai}
\affiliation{Department of Physics \& Astronomy, Rice Center for Quantum Materials,
Rice University, Houston, Texas 77005,USA}
\author{Rong Yu}
\email{rong.yu@ruc.edu.cn}
\affiliation{Department of Physics and Beijing Key Laboratory of Opto-electronic
Functional Materials and Micro-nano Devices, Renmin University of
China, Beijing 100872, China }
\author{Qimiao Si}
\email{qmsi@rice.edu}
\affiliation{Department of Physics \& Astronomy, Rice Center for Quantum Materials,
Rice University, Houston, Texas 77005,USA}
\date{\today}
\begin{abstract}
Elucidating the nature of spin excitations
is important
to understanding the mechanism of superconductivity
in the iron pnictides.
Motivated by recent inelastic neutron scattering measurements
in
the nearly 100\%
detwinned BaFe$_{2}$As$_{2}$, we study the spin dynamics of an
$S=1$ frustrated bilinear-biquadratic Heisenberg model in the antiferromagnetic
phase with wavevector $(\pi,0)$.  The biquadratic interactions are treated in a dynamical way using
a flavor-wave theory in an $SU(3)$ representation. Besides
the
dipolar spin wave (magnon)
excitations, the biquadratic interactions
give rise to quadrupolar excitations at high energies. We find that
the quadrupolar wave significantly influences, in an energy dependent way,
the
anisotropy between the
spin excitation
spectra
 along the $(\pi,0)$
and $(0,\pi)$ directions in the wave vector space. Our theoretical results
capture the essential behavior of the spin dynamics measured
in the antiferromagnetic phase of
the
detwinned BaFe$_2$As$_2$.
 More generally,
our results
underscore the
importance of
 electron correlation effects for the microscopic physics of the iron pnictides.
\end{abstract}
\maketitle

\section{Introduction}

\label{Sec:Intro} In iron pnictides, superconductivity develops near
an  antiferromagnetic order in the temperature-doping phase diagram.
It is therefore believed that understanding the nature of magnetic
excitations is crucial for uncovering the mechanism of superconductivity
in these materials.\cite{Kamihara_JACS:2008,Johnston_AP:2010,Wang_Sci:2011,Dai_RMP:2015,Si_NRM:2016,Hirschfeld_CRP:2016}
The iron pnictides are bad metals. Their room temperature resistivity,
about $0.4 \;$ m$\Omega$-cm, is larger than the Mott-Ioffe-Regel limit \cite{Johnston_AP:2010}.
This implicates strong electron-electron scatterings that are associated with the underlying electron correlations.
In addition, the optical conductivity reveals a large reduction of the Drude weight \cite{Qazilbash},
signifying a
 small coherent  electron spectral weight $w$; correspondingly,
the incoherent electron spectral weight ($1-w$)  is larger than the coherent electron counterpart.
To the zeroth order  in $w$, the system is located at the Mott transition; the entire single-electron excitations
are incoherent and they give rise to quasi-localized magnetic moments \cite{SiAbrahams}.
The coherent itinerant electrons with
weight $w$ will influence the spin excitation spectrum
at the linear and higher orders in $w$. \cite{SiAbrahams,SiNJP,DaiPNAS}
The interactions between the local moments include the $J_1$-$J_2$ Heisenberg
interactions. Moreover, in the regime near the Mott transition, the multiorbital nature
of the underlying electronic system implies that interactions
involving multiple spin operators such as
the biquadratic $K$ coupling naturally arise and can be sizeable \cite{Fazekas}.
A number of
additional
perspectives have been taken to consider the
electron correlation
effects
\cite{Fang:08,Xu:08,WChen,Moreo,Berg,Ma,MJHan,Laad.2009,Yin,Lv,Yu_PRB:2012,Wysocki,Uhrig,Ishida,Lorenzana,Bascones,Yu_PRL:2015,Wang_NatPhys:2015,Luo_PRB:2016,Goswami_PRB:2011,Stanek_PRB:2011,Yu_JPCS:2013,Ergueta_PRB:2015,Yu_PRL:2015,Wang_NatPhys:2015,Lai_PRL:2017,Ruiz_PRB:2019,Wang_arXiv:2019}.

The parent 122 iron pnictides, such as BaFe$_2$As$_2$,
exhibit a $(\pi,0)$ antiferromagnetic
(AFM) order at low temperatures.\cite{Dai_RMP:2015} Right at or slightly
above the magnetic ordering temperature $T_{N}$, a tetragonal-to-orthorhombic
structural transition takes place, breaking the $C_{4}$ spatial rotational symmetry.
As a consequence, the low-temperature spin excitations
are anisotropic in the wave vector
space
\cite{Zhao_NatPhys:2009,Inosov:2009,Lester:2010,Li:2010,Park:2010,Diallo_PRB:2010,Harriger_PRB:2011,Ewings_PRB:2011,LiuNatPhys12},
with equal-energy intensity distribution
forming ellipses
that are
centered around the wave vector $(\pi,0)$ at low energies.\cite{Diallo_PRB:2010,Harriger_PRB:2011}
With increasing energy, the elliptic feature expands and the spectral
weights transfer from $(\pi,0)$ to $(\pi,\pi)$ of the Brillouin
zone (BZ).\cite{Harriger_PRB:2011,Ewings_PRB:2011}

These properties are well understood by an effective $S=1$ bilinear-biquadratic
Heisenberg
model, for
the quasi-localized magnetic moments
that are
 produced by electron
correlations in bad metals such as the iron pnictides.\cite{Yu_PRB:2012,Goswami_PRB:2011,Stanek_PRB:2011,Luo_PRB:2016}
The Hamiltonian of this model reads as follows:
\begin{equation}
H=\frac{1}{2}\sum_{ij}\left\{ J_{ij}\mathbf{S}_{i}\cdot\mathbf{S}_{j}-K_{ij}(\mathbf{S}_{i}\cdot\mathbf{S}_{j})^{2}\right\} \label{eq:ham}
\end{equation}
where $J_{ij}=J_{1}$ and $K_{ij}=K_{1}$ are exchange couplings for
the nearest neighbor (NN) bonds on a square lattice, and $J_{ij}=J_{2}$
and $K_{ij}=K_{2}$ are for the next nearest neighbor (NNN) bonds.
In the remainder of this manuscript, we refer to the above model as the $J$-$K$
model.

Because the biquadratic term involves higher order spin interactions,
it is difficult to be handled within any conventional spin-wave theory.
For the iron pnictides,
the spin dynamics in its $(\pi,0)$ AFM phase
was thought to be described in terms of
an empirical $J_{1a}$-$J_{1b}$-$J_{2}$ model \cite{Zhao_NatPhys:2009,Applegate},
where $J_{1a}$ and $J_{1b}$ refer to the nearest-neighbor $J_1$ interactions
along the tetragonal $a$ and $b$ axes, respectively.
This effective description would arise from the $J$-$K$ model if one makes
a {\it static} approximation to the biquadratic interactions~\cite{Yu_PRB:2012},
with the biquadratic interaction being decomposed via a Hubbard-Stratonovich
field
 ($\Gamma_{ij}=\langle\mathbf{S}_{i}\cdot\mathbf{S}_{j}\rangle$)
 that
 is determined self-consistently.
However, the $J$-$K$ model is more fundamental than
the $J_{1a}$-$J_{1b}$-$J_{2}$ model. In contrast to the latter,
the $J$-$K$ model respects the underlying
tetragonal
 lattice symmetry.
For example, the $J$-$K$ model is able to describe the
spin excitation spectrum in the paramagnetic phase, in which $J_{1a}=J_{1b}$
is
dictated
by the tetragonal symmetry~\cite{Yu_PRB:2012}. It is worth noting that
the $J$-$K$ model has a very rich ground-state phase diagram.
In addition to the $(\pi,0)$ AFM order discussed here, it contains ground states
that explain the various emergent
nematicity in iron-based superconductors, such as in the bulk FeSe and
heavily hole-doped iron pnictides \cite{Yu_PRL:2015,Wang_arXiv:2019}.

Recent inelastic neutron scattering (INS) measurements on
almost fully
detwinned
BaFe$_{2}$As$_{2}$ samples~\cite{Lu_PRL:2018} provide new clues to
the nature of the spin excitations in these compounds.
Though the measured magnon
dispersion in the AFM phase can be well understood by
the  $J_{1a}$-$J_{1b}$-$J_{2}$ Heisenberg
model
up to about $100$ meV,
the anisotropy of the spectral weights between $(\pi,0)$ and $(0,\pi)$
in the Brillouin zone (BZ) can not. It
shows a strong energy dependence and,
contrary to the expectation of the $J_{1a}$-$J_{1b}$-$J_{2}$ model,
the observed
local ({\it i.e.}, ${\bf q}$-integrated)
spectral weights near $(\pi,0)$ and $(0,\pi)$ approach each other
 at high energies.\cite{Lu_PRL:2018}

Here we show that the seemingly unusual spin excitation anisotropy can be naturally understood in terms
of the dynamics induced by the biquadratic interaction. We start from the intuitive picture that the high-energy
spin excitations correspond to short-range and short-time fluctuations, whose spatial profile
will be similar to that of the paramagnetic phase; in the latter case, the spin excitations of the $J$-$K$ model
will be
$C_4$ symmetric,
which is to be contrasted with those of the
$J_{1a}$-$J_{1b}$-$J_{2}$ model
that
are inherently anisotropic.
With this picture in mind,
we will analyze the biquadratic $K$ interaction of the $S=1$ $J$-$K$ model
{\it dynamically}.
The ensuing
quadrupolar excitations at high energies
contribute to a
spin excitation spectrum
with a considerably reduced
anisotropy.
We calculate the dynamical spin susceptibilities from the microscopic model,
with
 results
 that
 semi-quantitatively account for
the puzzling neutron scattering results of the dewinned BaFe$_{2}$As$_{2}$.

More specifically, we analyze the model, Eq.~\eqref{eq:ham}, in terms of
an $SU(3)$
flavor-wave theory.
The
 $SU(3)$ representation~\cite{Mila},
which treats
the $J$ and $K$ terms
on an equal footing,
has previously been used in studying the spin dynamics
of iron-based superconductors \cite{Yu_PRL:2015,Luo_PRB:2016}.
In the AFM ground state, it incorporates the quadrupolar excitations along with
the magnetic dipolar (magnon) ones. While single
quadrupolar excitations are orthogonal to the dipolar channel,
their convolution with the magnon excitations, for instance, does contribute to the
dipolar channel and, hence, to the dynamical spin susceptibility.
In this way, the high-energy quadrupolar excitations significantly reduce the anisotropy
of the the high-energy spin excitation
spectrum; we
find that
the $J$-$K$ model provides an excellent understanding of the inelastic neutron
scattering experiments
in the detwinned BaFe$_{2}$As$_{2}$,
\cite{Lu_PRL:2018}
on
both
the
spin excitation anisotropy and
\emph{the spin spectral weights}.

We stress that, how to theoretically describe the spin excitations in iron-based superconductors
is an outstanding question of the field. The new experiments \cite{Lu_PRL:2018} on the spin excitation anisotropy
in the detwinned BaFe$_{2}$As$_{2}$ are particularly significant because they access the entire
magnetic band \cite{Harriger_PRB:2011}. Our analysis of this new experiment
allowed us to extract the effect of
quadrupolar
excitations in this canonical iron-pnictide system. For the iron chalcogenide FeSe, the role of antiferrroquadrupolar
channel had already been emphasized \cite{Yu_PRL:2015}. In this sense,
our work represents not only an advance for the description of the iron pnictides but also a new way
of unifying the overall understandings of both the iron pnictides and iron chalcogenides.

The remainder of the
paper
 is organized as follows.
In Sec.~\ref{Sec:Model}, we introduce
the $SU(3)$ representation for the $S=1$ $J$-$K$ model and describe the
calculation of the spin excitation spectrum within the $SU(3)$ flavor-wave
theory
\cite{Yu_PRL:2015,Luo_PRB:2016}.
We then present our main results, in
Sec.~\ref{Sec:Spectrum},
on the magnetic excitations of the $J$-$K$ model. We show how the
quadrupolar wave affects the anisotropy of the
spin
excitation spectrum
at high energies, and discuss in detail the frequency dependence of the spin spectral weights.
Both are shown to describe well the inelastic neutron scattering measurements
in the AFM phase of the detwinned
BaFe$_{2}$As$_{2}$.
In
Sec.~\ref{Sec:Discussion},
we contrast our results with those of weak-coupling analyses, describe the underestimation of the spin spectral weights
in an RPA calculation and discuss its implications,
before concluding the paper in Sec.~\ref{Sec:Conclusion}.

\section{Model and Method}

\label{Sec:Model} We start from the $S=1$ $J$-$K$
model defined in Eq.~\eqref{eq:ham}. For BaFe$_{2}$As$_{2}$,
the ground-state magnetic structure is a $\mathbf{Q}=(\pi,0)$ collinear
AFM order. We assume that the corresponding classical spin configuration
of this ordered state has all spins aligned in parallel along the
$S^{z}$ direction \emph{in the spin space}. To simplify the calculation
of the spin excitations, we first perform a site-dependent spin
rotation about the $y$-axis in the spin space:
\begin{equation}
\tilde{\mathbf{S}}_{i}=\hat{R}_{y}(\mathbf{Q}\cdot\mathbf{R}_{i})\mathbf{S}_{i}.
\end{equation}
After this rotation, the spins in even columns stay unchanged while
those in odd columns are rotated by a $\pi$ angle about the $y$
axis. Therefore, in the rotated configuration, all spins align ferromagnetically
along the (negative) $\tilde{S}^{z}$ direction, and the Hamiltonian
in the rotated basis keeps translational symmetry. We introduce
the $SU(3)$ flavor-wave representation for the (rotated) spin operators.
This is formally done by rewriting the spin operators in terms of
three flavor boson operators \cite{Yu_PRL:2015,Luo_PRB:2016}:
\begin{align}
\tilde{S}_{i}^{+} & =\sqrt{2}(b_{i1}^{\dagger}b_{i0}+b_{i0}^{\dagger}b_{i\overline{1}}),\label{eq:sbmap1}\\
\tilde{S}_{i}^{-} & =\sqrt{2}(b_{i\overline{1}}^{\dagger}b_{i0}+b_{i0}^{\dagger}b_{i1}),\label{eq:sbmap2}\\
\tilde{S}_{i}^{z} & =b_{i1}^{\dagger}b_{i1}-b_{i\overline{1}}^{\dagger}b_{i\overline{1}},\label{eq:sbmap3}
\end{align}
where $b_{i\alpha}^{\dagger}$ ($\alpha=1,0,\overline{1}$) creates
 a boson of flavor $\alpha$ on site $i$. The Hilbert
space of the bosons is larger than the original spin Hilbert
space and includes
unphysical states. To limit the boson Hilbert space to its physical
sector, a hard constraint is imposed on each site
\begin{equation}
b_{i1}^{\dagger}b_{i1}+b_{i0}^{\dagger}b_{i0}+b_{i\overline{1}}^{\dagger}b_{i\overline{1}}=1.\label{Eq:constraint}
\end{equation}

The magnetic ground state in the rotated spin space is polarized. In the bosonic
representation this corresponds to condensation of the $b_{\overline{1}}$
boson on each site, with the condensate amplitude
$\langle b_{i\overline{1}}\rangle$.
Note that the condensation in the bosonic representation depends
on the ground state property in the spin representation. For example,
for a quadrupolar ordered state that preserves time-reversal symmetry,
the condensation takes place in $b_{x}=(b_{1}+b_{\overline{1}})/\sqrt{2}$.

With
the
 $b_{\overline{1}}$ boson
 being condensed
 in the AFM ground state,
we
turn the constraint in Eq.~\eqref{Eq:constraint}
into the following:
\begin{equation}
b_{i\overline{1}}\approx\sqrt{1-b_{i1}^{\dagger}b_{i1}-b_{i0}^{\dagger}b_{i0}}.%
\label{Eq:constraint_mod}
\end{equation}
Deep in the AFM phase $b_{i\overline{1}}\sim O(1)$ and we can treat
$b_{i0}$ and $b_{i1}$ as perturbations. Rewriting the spin Hamiltonian
of Eq.~\eqref{eq:ham} in terms of the flavor bosons using Eqs.~\eqref{eq:sbmap1}-\eqref{eq:sbmap3},
and
expanding
 it in terms of $b_{i0}$ and $b_{i1}$ up to the quadratic
order using Eq.~\eqref{Eq:constraint_mod}, we obtain
\begin{align}
H\approx H_{2}=\frac{1}{2}\sum_{\mathbf{k}}\sum_{\nu=0,1}\left[A_{\mathbf{k}\nu}(b_{\mathbf{k}\nu}^{\dagger}b_{\mathbf{k}\nu}+b_{-\mathbf{k}\nu}^{\dagger}b_{-\mathbf{k}\nu})\right.\nonumber \\
\left.+B_{\mathbf{k}\nu}(b_{\mathbf{k}\nu}^{\dagger}b_{-\mathbf{k}\nu}^{\dagger}+b_{\mathbf{k}\nu}b_{-\mathbf{k}\nu})\right],%
\label{eq:H2}
\end{align}
where
\begin{align}
A_{\mathbf{k}0} & =2J_{1}\cos k_{y}+2K_{1}+4(J_{2}+K_{2}),\label{eq:a0}\\
B_{\mathbf{k}0} & =-2(J_{1}+K_{1})\cos k_{x} \nonumber\\
&~~~\, -4(J_{2}+K_{2})\cos k_{x}\cos k_{y},\label{eq:b0}\\
A_{\mathbf{k}1} & =8J_{2}-2K_{1}\cos k_{y}+4K_{2},\label{eq:a1}\\
B_{\mathbf{k}1} & =-2K_{1}\cos k_{x}-4K_{2}\cos k_{x}\cos k_{y}.\label{eq:b1}
\end{align}
With the $b_{\overline{1}}$ boson condensed, $b_{0}^{\dagger}$ corresponds
to creating a magnon (spin-1 dipolar excitation) that increases the
spin angular momentum $\tilde{S}^{z}$ by $1$, whereas $b_{1}^{\dagger}$
corresponds to creating spin-2 quadrupolar excitation that increases
 $\tilde{S}^{z}$ by $2$. The dipolar and quadrupolar operators do not
mix at the quadratic order because they respectively carry spin-$1$ and spin-$2$ angular momenta
(and, in addition, they possess different parities \cite{footnote}).

The quadratic Hamiltonian in Eq.~\eqref{eq:H2} can be diagonalized
via a Bogoliubov transformation
\begin{equation}
b_{k\nu}=u_{\mathbf{k}\nu}\beta_{\mathbf{k}\nu}+v_{\mathbf{k}\nu}\beta_{\mathbf{-k}\nu}^{\dagger}%
\end{equation}
where
\begin{align}
u_{\mathbf{k}\nu}=\sqrt{\frac{A_{\mathbf{k}\nu}+\omega_{\mathbf{k}\nu}}{2\omega_{\mathbf{k}\nu}}},\label{eq:uk}\\
v_{\mathbf{k}\nu}=-\mathrm{sgn}(B_{\mathbf{k}\nu})\sqrt{\frac{A_{\mathbf{k}\nu}-\omega_{\mathbf{k}\nu}}{2\omega_{\mathbf{k}\nu}}},\label{eq:vk}\\
\omega_{\mathbf{k}\nu}=\sqrt{A_{\mathbf{k}\nu}^{2}-B_{\mathbf{k}\nu}^{2}}.\label{eq:wk}
\end{align}
The diagonalized Hamiltonian reads
\begin{equation}
H_{2}=\sum_{\mathbf{k}}\sum_{\nu=0,1}\omega_{\mathbf{k}\nu}\beta_{\mathbf{k}\nu}^{\dagger}\beta_{\mathbf{k}\nu}+C,%
\label{eq:H2_Diag}
\end{equation}
where $C$ refers to the zero point energy of the AFM ordered state,
and $\beta_{\mathbf{k}0}$ and $\beta_{\mathbf{k}1}$ terms describe
the excitations of dipolar spin waves (magnons) and quadrupolar waves, respectively.

With the diagonalized Hamiltonian we can readily calculate the dynamical
structure factor (DSF) of spins, which is defined as
\begin{equation}
\mathcal{S}(\mathbf{q},\omega)=\int_{-\infty}^{\infty}\frac{\text{d}t}{2\pi}e^{i\omega t}\langle\mathbf{S}_{\mathbf{q}}(t)\cdot\mathbf{S}_{-\mathbf{q}}(0)\rangle,%
\label{eq:DSF}
\end{equation}
where $\mathbf{S}_{\mathbf{q}}=\frac{1}{\sqrt{N}}\sum_{\mathbf{k}}\mathbf{S}_{i}e^{i\mathbf{R}_{i}\cdot\mathbf{q}}$
is the Fourier transformed spin component.

\begin{figure}[t!]
\includegraphics[width=1\columnwidth]{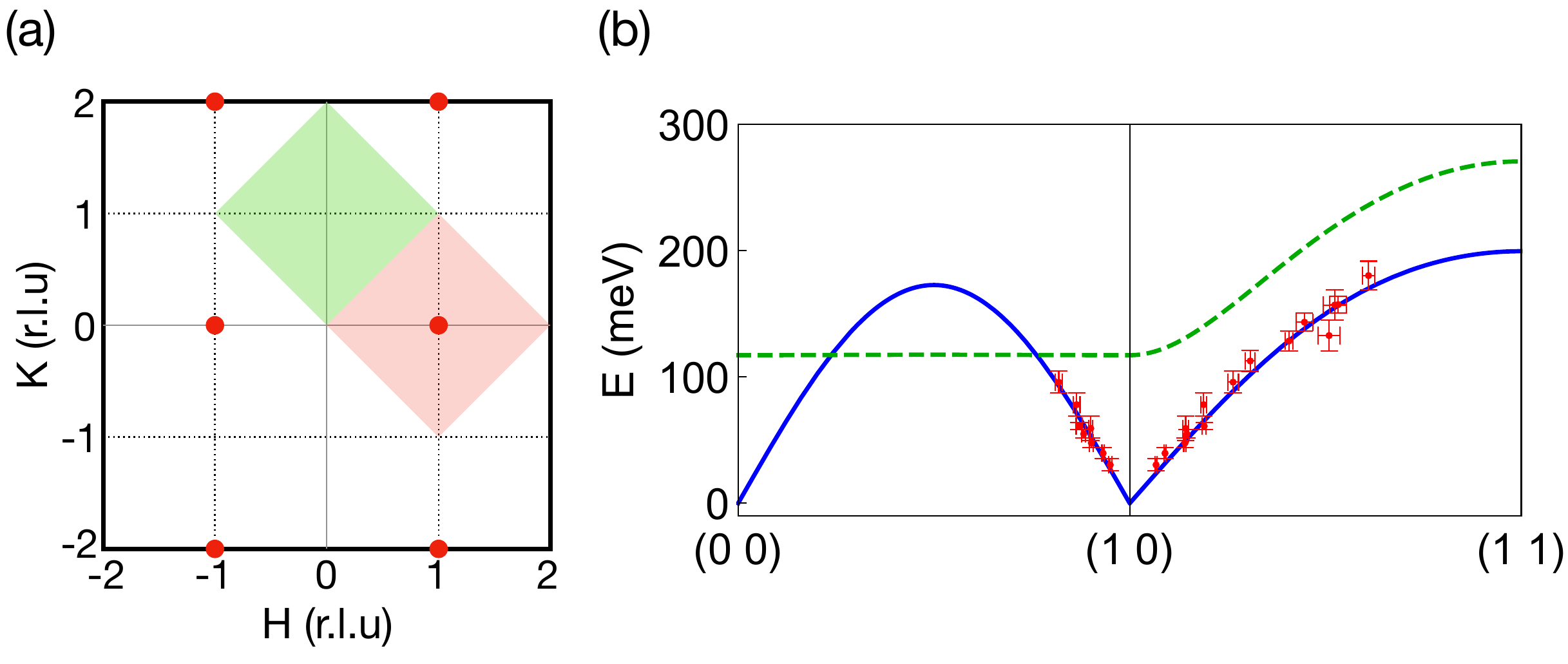} \caption{(a):
Reciprocal space of the detwinned BaFe$_{2}$As$_{2}$
in the $(\pi,0)$ AFM phase. The magnetic Bragg
peak positions of the $(\pi,0)$ magnetic order are marked as red
dots. The integrated areas centered at $(\pi,0)$ and $(0,\pi)$ in Eq.~\eqref{eq:chi} are shown by red and
green diamonds, respectively. (b): Dispersions of magnons (solid line) and quadrupolar
wave (dashed line) of the bilinear-biquadratic model from the $SU(3)$
flavor-wave theory. See text for the model parameters. The dots with
error bars show the measured dispersion data of a detwinned BaFe$_{2}$As$_{2}$
at $T=7$ K, extracted from Ref.~\onlinecite{Lu_PRL:2018}.}
\label{fig:dispersion}
\end{figure}

In the $SU(3)$ flavor-wave theory the DSF can be separated into two parts,
$\mathcal{S}(\mathbf{q},\omega)=\mathcal{S}_{c}(\mathbf{q},\omega)+\mathcal{S}_{i}(\mathbf{q},\omega)$.
The coherent part $\mathcal{S}_{c}(\mathbf{q},\omega)$ comes from
one-magnon process
\begin{equation}
\mathcal{S}_{c}(\mathbf{q},\omega)=(u_{\mathbf{q}0}-v_{\mathbf{q}0})^{2}\delta(\omega-\omega_{\mathbf{q}0}),%
\label{eq:Sc}
\end{equation}
whereas the incoherent part $\mathcal{S}_{i}(\mathbf{q},\omega)$
contains various two-particle contributions.
\begin{align}
 & \mathcal{S}_{i}(\mathbf{q},\omega)\nonumber \\
 & =\frac{1}{N}\sum_{\mathbf{k},\mathbf{k}'=\mathbf{q}-\mathbf{k}}\Big[(v_{\mathbf{k}0}u_{\mathbf{k}'1}-u_{\mathbf{k}0}v_{\mathbf{k}'1})^{2}\delta(\omega-\omega_{\mathbf{k}1}-\omega_{\mathbf{k}'0})\nonumber \\
 & +2(u_{\mathbf{k}1}v_{\mathbf{k}'1}-v_{\mathbf{k}1}u_{\mathbf{k}'1})^{2}\delta(\omega-\omega_{\mathbf{k}1}-\omega_{\mathbf{k}'1})\nonumber \\
 & +\frac{1}{2}(u_{\mathbf{k}0}v_{\mathbf{k}'0}-v_{\mathbf{k}0}u_{\mathbf{k}'0})^{2}\delta(\omega-\omega_{\mathbf{k}0}-\omega_{\mathbf{k}'0})\Big].\label{eq:Si}
\end{align}
where the three terms from top to bottom on the right hand side correspond
to contributions from magnon-quadrupole, two-quadrupole, and two-magnon
processes, respectively \cite{Luo_PRB:2016}.

\section{Spin excitation spectrum}

\label{Sec:Spectrum}

\subsection{Magnon and quadrupolar-wave dispersions}

\begin{figure*}[t!]
\includegraphics[width=0.9\textwidth]{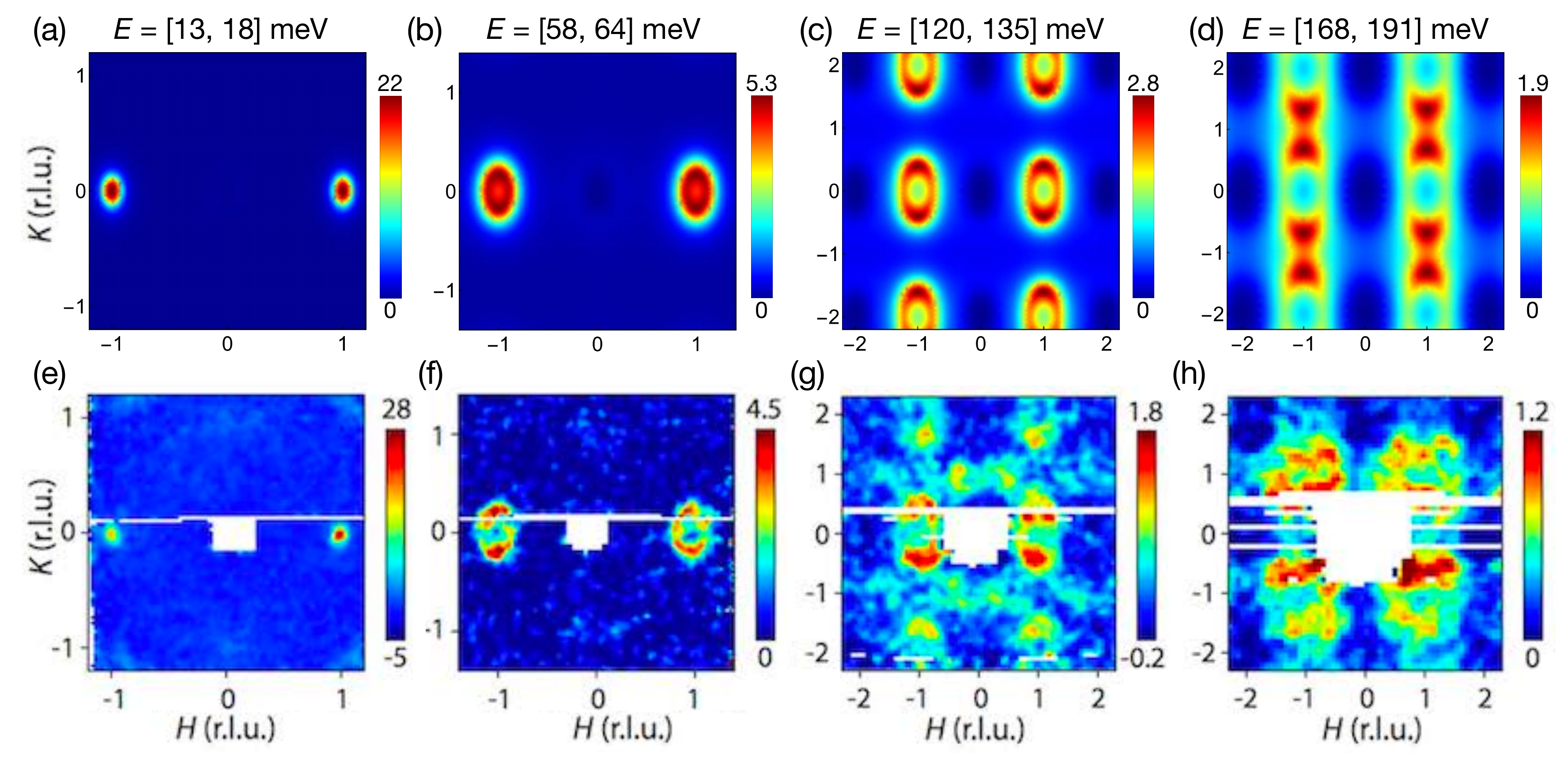} \caption{Constant energy cuts of the magnetic excitation spectrum
in the wave vector space
for the $(\pi,0)$ AFM phase of the $J$-$K$ model {[}in (a)-(d){]}.
For comparison, the corresponding experimental results are shown in
(e)-(h). The experimental plots are reproduced from Ref.~\onlinecite{Lu_PRL:2018}.}
\label{fig:intensity}
\end{figure*}

To understand the spin excitations in iron pnictides we start
by calculating the spin-wave (magnon) and quadrupolar-wave dispersions
in the $J$-$K$ model from Eqs.~\eqref{eq:H2}-\eqref{eq:H2_Diag}.
The result is shown in Fig.~\ref{fig:dispersion}. The model parameters
used in this plot are $J_{1}=-9.2\pm1.2$ meV, $J_{2}=50.0\pm5.0$
meV, $K_{1}=68.4\pm3.2$ meV, and $K_{2}=-36.4\pm5.0$ meV.
They allow for a good fit
to the experimentally observed
magnon dispersion and,
as we see below, define a model
whose solution provides a semi-quantitative understanding of the
experimentally measured
dynamical
spin susceptibility.
In Fig.~\ref{fig:dispersion},
the
solid line shows the magnon dispersion $\omega_{\mathbf{k}0}$.
It clearly
characterizes
 the Goldstone mode near $(\pi,0)$ and a flat band
top
near
$(\pi,\pi)$. The dispersion agrees well with
that measured in the detwinned
BaFe$_{2}$As$_{2}$, which is described by the
symbols in the figure~\cite{Harriger_PRB:2011,Lu_PRL:2018}.
In addition to the magnon branch, there is a quadrupolar-wave
branch in the $SU(3)$ representation, whose dispersion is shown as
the dashed line in Fig.~\ref{fig:dispersion}. The quadrupolar excitation
carries spin angular momentum 2, and can be viewed as a two-magnon
bound state. As such, it is generically gapped in an AFM ordered
phase. As shown in Fig.~\ref{fig:dispersion}, for BaFe$_{2}$As$_{2}$
the quadrupolar-wave excitation gap is
about
100 meV. We then expect
it to primarily influence
the spin excitation
spectrum
at
energies
$\gtrsim100$ meV.
 This is the
reason why the spectrum below about 100 meV (where the magnon dispersion
can be probed) can be well understood by the spin-wave theory in the conventional
$SU(2)$ representation~\cite{Yu_PRB:2012}.

As seen in Eq.~\eqref{eq:Sc}, the quadrupolar excitations cannot be
directly detected by inelastic neutron measurements since quadrupoles
carry spin-2 that do not directly couple to neutrons; the coherent
part is only contributed by the one-magnon process. However, two quadrupoles
can form an effective spin-1 object that transforms as a magnetic
dipole moment under $SU(2)$ spin rotation operation; correspondingly,
the two-quadrupole
processes do contribute to the dynamical spin susceptibility.
From similar considerations, the same applies to
the one-quadrupole-one-magnon processes. Therefore, quadrupole
excitations are manifested as the incoherent continuum of the spin excitation spectrum
and are
expected to appear at high energies in the inelastic neutron scattering spectrum.

\subsection{Dynamical structure factor}
We have calculated the spin DSF of the $J$-$K$ model
using Eqs.~\eqref{eq:DSF}-\eqref{eq:Si} within the $SU(3)$ flavor-wave theory.
The coupling of the quasi-localized magnetic moments with coherent electrons
will produce nonzero damping rates~\cite{Yu_PRB:2012,Goswami_PRB:2011},
 which will broaden
 the $\delta$ functions
 into
 a damped harmonic oscillator profile~\cite{Harriger_PRB:2011},
\begin{equation}
\delta(\omega-\omega_{\mathbf{q}0})\sim\frac{4}{\pi}\frac{\Gamma_{\mathbf{q}}\omega\omega_{\mathbf{q}0}}{(\omega^{2}-\omega_{\mathbf{q}0}^{2})^{2}+4(\Gamma_{\mathbf{q}}\omega)^{2}},%
\label{eq:delta_func}
\end{equation}
where
$\Gamma_{\mathbf{q}}=\Gamma_{0}+A\cos^{2}\frac{q_{x}}{2}+B\cos^{2}\frac{q_{y}}{2}$.
The damping parameters are taken from Ref.~\onlinecite{Harriger_PRB:2011}.

We show the constant energy cuts of the calculated DSF at several excitation
energies in Fig.~\ref{fig:intensity}(a)-(d).
At low energies, the peaks of DSF form an ellipse centered at $(\pi,0)$
in the first BZ. This is consistent with the $(\pi,0)$ AFM ground
state of the system, above which the low-energy excitations are the
Goldstone modes around the ordering wave vector $(\pi,0)$. By contrast,
the spin excitations
around $(0,\pi)$  are gapped, reflecting the broken
$C_{4}$ symmetry of the ground state. With increasing energy the
ellipse centered at $(\pi,0)$ expands as shown in Fig.~\ref{fig:intensity}(a)
and (b). Further increasing the energy above 100 meV, going along
the ellipse the spectral weight in the $q_{x}$ direction largely
decreases and the maximum of the spectral weight is distributed at
the long-axis ($q_{y}$) direction. The ellipse then effectively splits
into two parts and the spectral weights transfer along the $q_{y}$
direction towards to $(\pi,\pi)$, as shown in Fig.~\ref{fig:intensity}(c)
and (d). These
results capture the main features of
the measured DSF
of detwinned
BaFe$_{2}$As$_{2}$ in the corresponding energies shown in Fig.~\ref{fig:intensity}(e)-(h).\cite{Lu_PRL:2018}

\begin{figure}[t!]
\includegraphics[width=0.8\columnwidth]{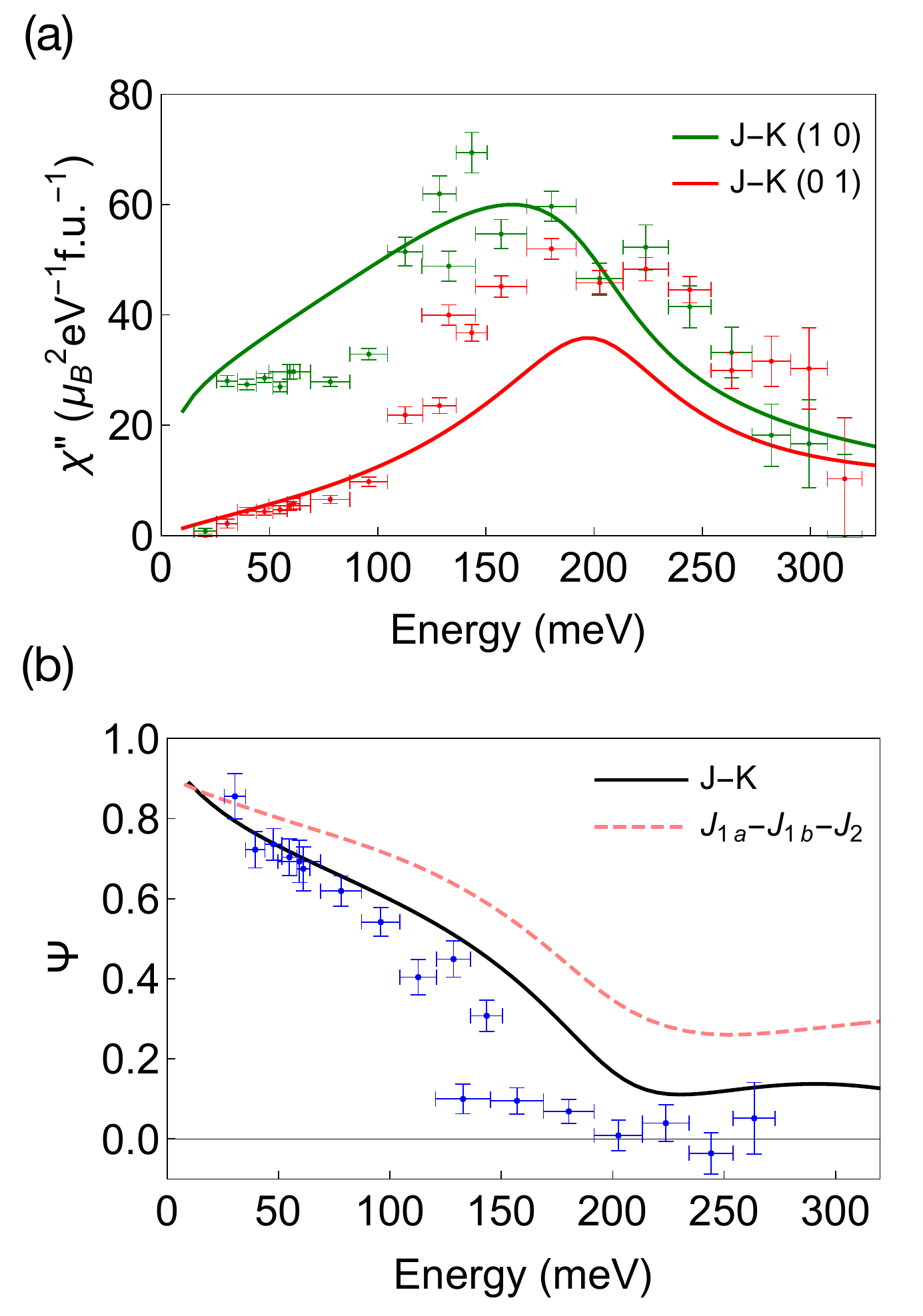}
\caption{
(a): Comparison of the energy dependence for the local dynamical spin susceptibilities
$\chi_{\mathbf{Q}_{1(2)}}$, as defined in Eq.~\ref{eq:chi}, with
$\mathbf{Q}_{1}=(\pi,0)$ and $\mathbf{Q}_{2}=(0,\pi)$ respectively marked
as $(1,0)$ and $(0,1)$,
 between
the
experimental data (symbols) and
the $SU(3)$ flavor-wave theory results of the $J$-$K$ model (lines). (b): Comparison of the spin excitation anisotropy factor $\Psi$
between the experimental data (symbols),
 the $SU(3)$ flavor-wave results of the $J$-$K$ model (black solid curve),
 and the $SU(2)$ spin-wave results of the $J_{1a}$-$J_{1b}$-$J_2$ model (pink solid curve).
 The experimental data for $\chi_{\mathbf{Q}_{1(2)}}$ in (a)
 are from Ref.~\onlinecite{Lu_PRL:2018}, from which
the experimental data for $\Psi$ in (b) are determined according to
Eq.~\eqref{Psi:def}.
The result (pink solid curve)
for $\Psi$ from an $SU(2)$-based calculation of the $J_{1a}$-$J_{1b}$-$J_2$ model,
shown in (b), is also from Ref.~\onlinecite{Lu_PRL:2018}.
}
\label{fig:anisotropy}
\end{figure}

The
DSF calculated from the $SU(3)$ flavor-wave
theory also shows some interesting characteristics that are not captured
by the conventional spin-wave theory. In particular, the spectral
weights close to $(\pi,0)$ and $(0,\pi)$ are very different at low
energies, but this anisotropy is reduced with increasing energy.
To see this clearly we calculate the local dynamical susceptibilities
$\chi_{\mathbf{Q}_{i}}^{\prime\prime}(\omega)$ over the wave vector regimes
near $\mathbf{Q}_{1}=(\pi,0)$ and $\mathbf{Q}_{2}=(0,\pi)$, respectively,
\begin{equation}
\chi_{\mathbf{Q}_{1(2)}}^{\prime\prime}(\omega)=\frac{\int_{\mathbf{q}\in{\rm {BZ}_{\mathbf{Q}_{1(2)}}}}d\mathbf{q}\chi_{\mathbf{q}}^{\prime\prime}(\omega)}{\int_{\mathbf{q}\in{\rm {BZ}_{\mathbf{Q}_{1(2)}}}}d\mathbf{q}}.
\label{eq:chi}
\end{equation}
where the integration regime ${\rm {BZ}_{\mathbf{Q}_{1(2)}}}$ is
shown in Fig.~\ref{fig:dispersion}(a),
as red and green diamonds respectively.
An anisotropy factor is defined as
\begin{equation}
\Psi(\omega)=\frac{\chi_{\mathbf{Q}_{1}}^{\prime\prime}(\omega)-\chi_{\mathbf{Q}_{2}}^{\prime\prime}(\omega)}{\chi_{\mathbf{Q}_{1}}^{\prime\prime}(\omega)+\chi_{\mathbf{Q}_{2}}^{\prime\prime}(\omega)}.
\label{Psi:def}
\end{equation}
The energy dependence of the local susceptibilities and the anisotropy
factor $\Psi$ are shown in Fig.~\ref{fig:anisotropy}. Both local
susceptibilities develop broad peaks in between 150 and 200 meV, and
exhibit considerable spectral weights up to 300 meV.
A major contribution to the high-energy spectral weights
comes from the incoherent part of the DSF. The difference
between the two local susceptibilities
is reduced
with increasing energy,
and this is clearly seen from the decreasing anisotropy factor $\Psi$
with increasing energy, as shown in Fig.~\ref{fig:anisotropy}(b). Such a strong
energy dependent spectral weight anisotropy is also observed in the
neutron scattering experiment, and our theoretical results agree well with the
experimental data. However, as shown in Fig.~\ref{fig:anisotropy}(b), this feature is not captured
by the effective $J_{1a}$-$J_{1b}$-$J_{2}$ model, in which
 the anisotropy as calculated from the conventional
spin-wave theory
persists to high energies owing to
its intrinsic anisotropic nature.\cite{Lu_PRL:2018}

To understand this
energy-dependent spectral weight anisotropy,
we recall that the high-energy spectrum
contains or is even dominated by the contribution from incoherent
part of the DSF. The latter contains various two-particle
processes. For a given set of $(\mathbf{q},\omega)$ there can be
many two-particle processes satisfying the energy and momentum conservation,
as shown by the additional summation over $\mathbf{k}$ in Eq.~\eqref{eq:Si}.
Some processes may contribute equally to the local spectral weights
near $(\pi,0)$ and $(0,\pi)$ and hence reduce the spectral anisotropy.
For example, for a given $\omega$ the contribution to $\mathcal{S}_{i}(\pi,0)$
may come from quasiparticles with momenta $(\pi/2,\pi/2)$ and $(\pi/2,-\pi/2)$,
while the contribution to $\mathcal{S}_{i}(0,\pi)$ may come from
quasiparticles with momenta $(\pi/2,\pi/2)$ and $(-\pi/2,\pi/2)$.
Since $(\pm\pi/2,\pm\pi/2)$ are equivalent points in the BZ, their
contributions to $\mathcal{S}_{i}(\pi,0)$ and $\mathcal{S}_{i}(0,\pi)$
are equal.
While all the three types of two-particle processes
in Eq.~\eqref{eq:Si} in principle contribute to
the reduction of spectral anisotropy,
the dominant contribution at high energies
involves the quadrupolar excitations.

\begin{figure}[t!]
\includegraphics[width=0.8\columnwidth]{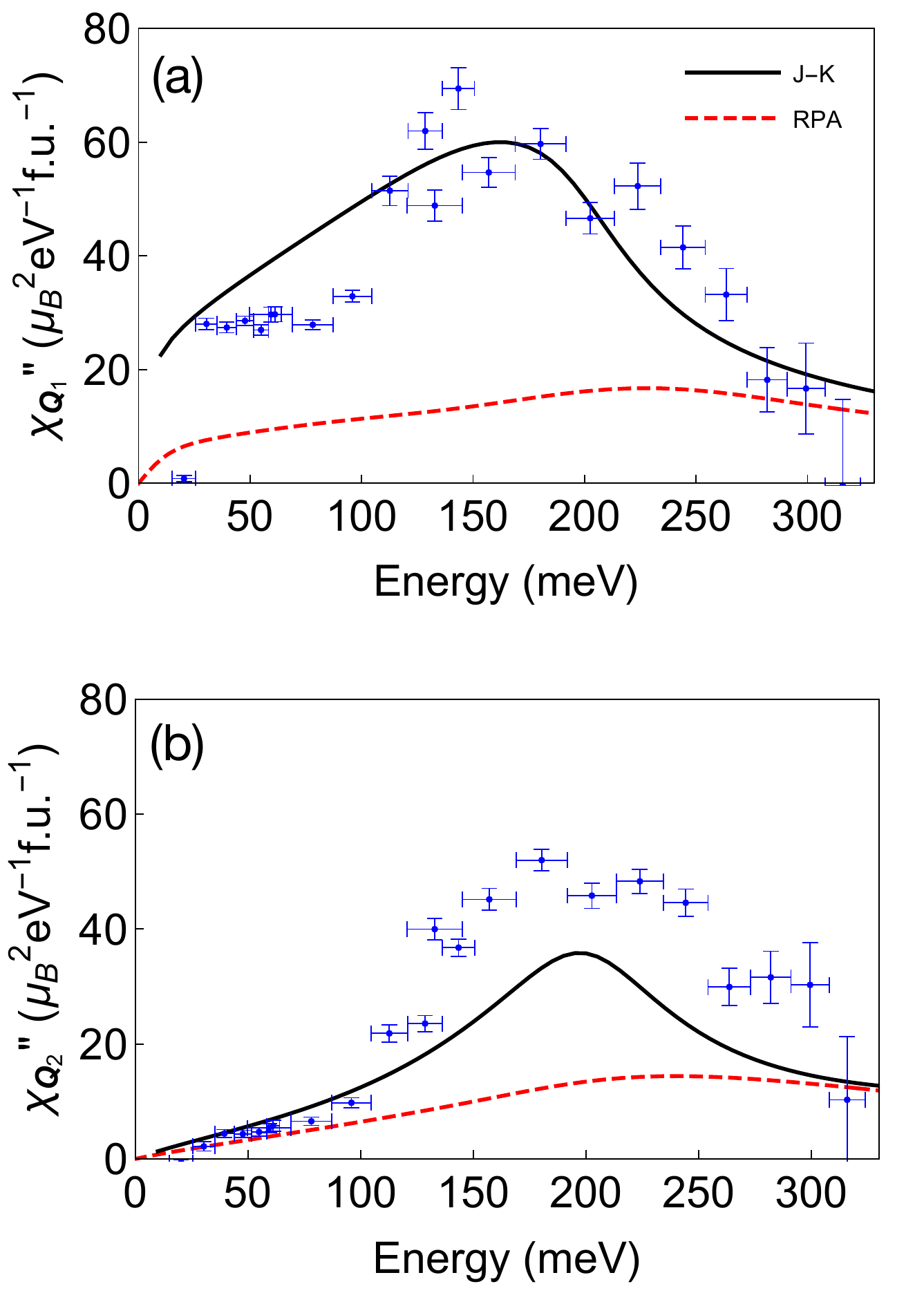} \caption{The energy dependence
of the local susceptibilities $\chi^{\prime\prime}$:
comparing the experimental data (symbols),
the $SU(3)$ flavor-wave theory of the $J$-$K$ model (solid lines),
and the modified RPA results (dashed lines) in absolute units.
The experimental data and the modified RPA results are
from Ref.~\onlinecite{Lu_PRL:2018}.}
\label{fig:appendix}
\end{figure}

\section{Discussions}
\label{Sec:Discussion}

In this work, we have provided a semi-quantitative understanding of the anisotropic spin excitation spectrum
observed by inelastic neutron scattering experiments in the detwinned BaFe$_{2}$As$_{2}$.
Treating the correlation-induced incoherent electronic excitations in terms of quasi-local moments,
we determined the anisotropic dynamical spin susceptibilities of
 the spin $S=1$ $J$-$K$ model. We did so by analyzing the biquadratic $K$ interaction
  dynamically, based on an $SU(3)$ representation of the spin.
Our results, for both the momentum distribution of the dynamical spin susceptibility and
the spin excitation anisotropy factor $\Psi$, are consistent with the experimental results measured in the
detwinned
 BaFe$_{2}$As$_{2}$.\cite{Lu_PRL:2018}

The experimental results have alternatively been analyzed
in terms of
electron-hole excitations within a modified random-phase approximation (RPA) calculation~\cite{Lu_PRL:2018}.
Weak-coupling approaches, such as
RPA calculations, represent an alternative means
to realize a $(\pi,0)$ AFM phase in the parent iron pnictides.\cite{Graser,Ran,Knolle}
Given the relatively small size of the Fermi pockets in
these systems,
 the standard RPA calculation was known to produce
too small
a spin spectral weight
compared to
 the experimental measurement.
The modified RPA calculation tried to remedy this by introducing a quasiparticle weight
$z<1$ to mimic the interaction effects. The result of this modified RPA calculation still under-accounts for the experimentally
measured dynamical spin susceptibility. This is illustrated
in Fig.~\ref{fig:appendix}, where we compare the experimental spectral weight data (and our results for the $J$-$K$ model)
with those from the modified RPA calculation in absolute units.
The
substantial under-estimation
of  the spin spectral weight even in the modified RPA calculation points to the dominating contributions from the incoherent electron excitations,
which are not captured by such calculations.

In our approach, we account for the spin excitations derived from the incoherent part of the single-electron excitations through quasi-localized magnetic moments and describe them in terms of the $J$-$K$ model.
The semi-quantitative success of our calculations
in capturing
 the experimentally measured spin excitation spectrum
 reinforces the above conclusion.
It suggests that the overall spin dynamics of the iron pnictides in an extended energy range
 is well described by approaches that are anchored by the fluctuations of local moments.
 Because the latter describes the spin degrees of freedom produced by the correlation-driven incoherent
 electronic excitations,
 our results also imply that the energy scales associated with the anisotropic magnetic fluctuations
 will be very large as has indeed been observed experimentally (and failed to be described by the modified RPA calculations)
 \cite{Lu_PRL:2018}.
 All these aspects underscore
the importance of electron correlations in the iron pnictides.

 The above considerations
 reinforce the implications that have been
 drawn from the presence of Mott insulating phases in
 both the iron chalcogenide \cite{Zhu10,Free10,Freelon15,MFang,DMWang,Wang_PRB:2015}
 and iron pnictide \cite{Song_NatComm:2016} families.
 An exciting recent development in the same spirit is the observation of magnetic and nematic orders
 in a semiconducting iron chalcogenide
 KFe$_{0.8}$Ag$_{1.2}$Te$_2$; \cite{Song_PRL:2019}
 exploration of the presumably anisotropic spin dynamics
 in this and related systems promises to shed further light on the role of electron correlations in the overall
 physics of both the iron pnictides and chalcogenides.

We close this section by noting on several additional points. Firstly, as we alluded to in the introduction,the discrepancy of the observed spin excitation anisotropy with the description by the $J_{1a}$-$J_{1b}$-$J_{2}$ model is intrinsic and, in particular,
is not a reflection of any particular parameter choice. The large $J_{1a}$-$J_{1b}$ anisotropy that is needed to
understand the spin-wave dispersion inherently
implies a large spin excitation anisotropy even for the high energy
magnetic excitations at the magnetic zone boundary, in contrast to the experimental observation.

Secondly, and in contrast to the $J_{1a}$-$J_{1b}$-$J_2$ description,
the Hamiltonian of the bilinear-biquadratic $J$-$K$
model itself respects the tetragonal symmetry.
Even with the spontaneous symmetry breaking, in the (pi,0) antiferromagnetic phase,
the emergence of quadrupolar excitations
hastens the restoration of the tetragonal symmetry at high energies,
precisely as the experimental results show.
In this sense, the measurement of the high-energy spin excitation anisotropy under the detwinning condition
is currently the most direct way of probing the quadrupolar excitations in the iron pnictides.

Our theory
implicates
 a quadrupolar wave excitation existing above 100 meV, though it is hidden to neutron scattering measurements.
In principle, the quadrupolar and dipolar moments may mix under a certain magnetic field~\cite{Smerald2013,Barzykin},
thereby activating the quadrupolar excitations in magnetic dipolar probes such as neutron scattering experiments.
This effect would be significant only when the Zeeman energy of the applied magnetic field is comparable to the
typical magnetic energy scale,
as happens in heavy fermion systems \cite{Portnichenko19, Paschen_2012}.
For the canonical iron pnictide system we discussed, however,
the magnetic bandwidth is larger than 100 meV, and the involved
exchange coupling is
on the order of
10 meV.
These energy scales are much
larger than
that associated with
the typical strength of the magnetic field applied in neutron scattering experiments.
Hence, within the available experimental capacities,
we do not expect the application of an external magnetic field to
help resolve the quadrupolar wave excitations more explicitly.

\section{Summary}
\label{Sec:Conclusion}

To summarize, we have investigated the spin excitations of an $S=1$
bilinear-biquadratic Heisenberg model in the antiferromagnetic phase.
We find that the magnetic excitations
consist of a dipolar spin wave (magnon) at low energies and a quadrupolar
wave at high energies. Though the quadrupolar excitations cannot
be directly detected by neutron scattering, we show that they can
significantly affect the spin excitation spectrum and, in particular,
reduce the
anisotropy
at high energies
 between the local susceptibilities near $(\pi,0)$ and $(0,\pi)$
in the wave vector space. Our theoretical results capture the essential
features of the spin excitations
of
the detwinned BaFe$_{2}$As$_{2}$. This suggests that
the incoherent part of the single-electron excitations, which give rise to quasilocalized magnetic moments,
dominates
 the contributions to the spin excitations.\cite{SiNJP,DaiPNAS}
Correspondingly, this
 implies that electron correlations play a central role
 in
 the microscopic physics of the
 iron-based superconductors.

\section*{Acknowledgement}

We thank the late
Elihu
 Abrahams, Haoyu Hu, Wenjun Hu and Hsin-Hua Lai for useful discussions.
This work
was supported
by the Ministry of Science and Technology of China Grants No. 2016YFA0301001, 2016YFA0300500,
and 2018YFGH000095 (C.L.),
 by the National Natural Science Foundation of China
Grant No. 11734002 (X.L.),
 by the Ministry of Science and Technology of China Grant No. 2016YFA0300504, the National Natural Science Foundation
 of China Grant No. 11674392, the Fundamental Research Funds for the Central Universities,
  the Research Funds of Renmin University of China Grant No. 18XNLG24 (R.Y.),
  and
by U. S. NSF DMR-1700081 (P.D.) and by the U.S. Department of Energy, Office of Science, Basic Energy Sciences,
under Award No. DE-SC0018197 and the Robert A. Welch Foundation Grant No. C-1411
 (Q.S.).


\begin{thebibliography}{10}
\bibitem{Kamihara_JACS:2008} Y. Kamihara, T. Watanabe, M. Hirano,
and H. Hosono, 
J. Am. Chem. Soc. \textbf{130}, 3296 (2008).

\bibitem{Johnston_AP:2010} D.~C. Johnston, 
Adv. Phys. \textbf{59}, 803-1061 (2010).

\bibitem{Wang_Sci:2011} F. Wang and D.-H. Lee, 
Science \textbf{332}, 200-204 (2011).

\bibitem{Dai_RMP:2015} P. Dai, 
Rev. Mod. Phys. \textbf{87}, 855-896 (2015).

\bibitem{Si_NRM:2016} Q. Si, R. Yu and E. Abrahams, Nat. Rev. Mater.
\textbf{1}, 16017 (2016). 

\bibitem{Hirschfeld_CRP:2016} P. J. Hirschfeld, 
Comptes Rendus Physique \textbf{17}, 197 (2016).

\bibitem{Qazilbash}  M. Qazilbash, J. Hamlin, R. Baumbach, L. Zhang, D. J. Singh, M. Maple, and D. Basov, 
Nat. Phys. \textbf{5}, 647 (2009).

\bibitem{SiAbrahams} Q. Si and E. Abrahams, 
Phys. Rev. Lett. \textbf{101}, 076401 (2008).

\bibitem{SiNJP}
Q. Si, E. Abrahams, J. Dai, and J.-X. Zhu,
New J. Phys. \textbf {11}, 045001 (2009).

\bibitem{DaiPNAS} J. Dai, Q. Si, J.-X. Zhu, and E. Abrahams,
Proc. Natl. Acad. Sci. \textbf{106}, 4118 (2009).

\bibitem{Fazekas} P. Fazekas, \emph{Lecture Notes on Electron Correlation
and Magnetism} (World Scientific, Singapore, 1999), Chap. 5.

\bibitem{Fang:08}
C. Fang, H. Yao, W.-F. Tsai, J. Hu and S. A. Kivelson
Phys. Rev. B \textbf{77}, 224509 (2008).

\bibitem{Xu:08}
C. Xu, M. Muller and S. Sachdev,
Phys. Rev. B \textbf{78}, 020501(R) (2008).

\bibitem{WChen} W.-Q. Chen, K.-Y. Yang, Y. Zhou, and F.-C. Zhang,
Phys. Rev. Lett. \textbf{102}, 047006 (2009).

\bibitem{Moreo} A. Moreo, M. Daghofer, J. A. Riera and E. Dagotto,
Phys. Rev. B \textbf{79}, 134502 (2009).

\bibitem{Berg}
E. Berg, S. A. Kivelson and D. J. Scalapino,
New J. Phys. \textbf{11}, 085007 (2009).

\bibitem{Ma}
F. Ma, Z.-Y. Lu, and T. Xiang,
Phys. Rev. B \textbf{78},  224517 (2008).

\bibitem{MJHan}
M. J. Han, Q. Yin, W. E. Pickett and S. Y. Savrasov,
Phys. Rev. Lett. \textbf{102} 107003 (2009).

\bibitem{Laad.2009} M. S. Laad, L. Craco, S. Leoni and H. Rosner,
Phys. Rev. B {\bf 79}, 024515 (2009).

\bibitem{Yin}
Z. P. Yin, K. Haule, and G. Kotliar,
Nature Mater. \textbf{10}, 932 (2011).

\bibitem{Lv}
W. Lv, F. Kr\"{u}ger and P. Phillips,
Phys. Rev. B \textbf{82}, 045125 (2010).

\bibitem{Yu_PRB:2012} R. Yu, Z. Wang, P. Goswami, A. H. Nevidomskyy,
Q. Si, and E. Abrahams, Phys. Rev. B \textbf{86}, 085148 (2012).

\bibitem{Wysocki}
A. L. Wysocki, K. D. Belashchenko and V. P. Antropov,
Nat. Phys. \textbf{7}, 485 (2011).

\bibitem{Uhrig}
G. S. Uhrig, M. Holt, J. Oitmaa, O. P. Sushkov and R. R. P. Singh,
Phys. Rev. B \textbf{79}, 092416 (2009).

\bibitem{Ishida}
H. Ishida and A. Liebsch,
Phys. Rev. B \textbf{81}, 054513 (2010).

\bibitem{Lorenzana}
G. Giavannetti, C. Ortix, M. Marsman, M. Capone,
J. van den Brink and J. Lorenzana,
Nat. Commun. \textbf{2}, 398 (2011).

\bibitem{Bascones}
M. J. Calder\'{o}n, F. Le'{o}n, B. Valenzuela and E. Bascones,
Phys. Rev. B \textbf{86}, 104514 (2012).

\bibitem{Yu_PRL:2015} R. Yu and Q. Si, Phys. Rev. Lett. \textbf{115},
116401 (2015).

\bibitem{Wang_NatPhys:2015}
F. Wang, S. A. Kivelson, and D.-H. Lee,
Nat. Phys. \textbf{11},959 (2015).

\bibitem{Luo_PRB:2016} C. Luo, T. Datta, and D.-X. Yao, Phys. Rev.
B \textbf{93}, 235148 (2016).

\bibitem{Goswami_PRB:2011}
P. Goswami, R. Yu, Q. Si, and E. Abrahams,
Phys. Rev. B {\bf 84}, 155108 (2011).

\bibitem{Stanek_PRB:2011} D. Stanek, O. P. Sushkov, and G. S. Uhrig,
Phys. Rev. B \textbf{84}, 064505 (2011).

\bibitem{Yu_JPCS:2013}
R. Yu, Q. Si, P. Goswami, and E. Abrahams,
J. Phys.: Conf. Ser. {\bf 449}, 012025 (2013).

\bibitem{Ergueta_PRB:2015}
P. B. Ergueta and A. H. Nevidomskyy,
Phys. Rev. B \textbf{92}, 165102 (2015).

\bibitem{Lai_PRL:2017}
H.-H. Lai, W.-J. Hu, E. M. Nica, R. Yu, and Q. Si, Phys. Rev. Lett. {\bf 118}, 176401 (2017).

\bibitem{Ruiz_PRB:2019}
H. Ruiz, Y. Wang,  B. Moritz, A. Baum, R. Hackl, and T. P. Devereaux,
Phys. Rev. B \textbf{83}, 214519 (2011).

\bibitem{Wang_arXiv:2019} Y. Wang, W. Hu, R. Yu, and Q. Si,
Phys. Rev. B \textbf{100},100502(R) (2019).


\bibitem{Zhao_NatPhys:2009} J. Zhao, D. Adroja, D.-X. Yao, R. Bewley, S. Li, X. Wang, G. Wu, X. Chen, J. Hu, and P. Dai, Nat. Phys. \textbf{5}, 555 (2009).

\bibitem{Inosov:2009} D. Inosov, J. Park, P. Bourges, D. Sun, Y. Sidis, A. Schneidewind, K. Hradil, D. Haug, C. Lin, B. Keimer, and V. Hinkov,  Nat. Phys. \textbf{6}, 178 (2010).

\bibitem{Lester:2010}
C. Lester, J.-H. Chu, J. G. Analytis, T. G. Perring, I. R. Fisher, and S. M. Hayden, Phys. Rev. B, \textbf{81}, 064505 (2010).

\bibitem{Li:2010}
H.-F. Li, C. Broholm, D. Vaknin, R. M. Fernandes, D. L. Abernathy, M. B. Stone, D. K. Pratt, W. Tian, Y. Qiu, N. Ni, S. O. Diallo, J. L. Zarestky, S. L. Bud'ko, P. C. Canfield, and R. J. McQueeney, Phys. Rev. B \textbf{82}, 140503 (2010).

\bibitem{Park:2010}
J. T. Park, D. S. Inosov, A. Yaresko, S. Graser, D. L. Sun, P. Bourges, Y. Sidis, Y. Li, J.-H. Kim, D. Haug, A. Ivanov, K. Hradil, A. Schneidewind, P. Link, E. Faulhaber, I. Glavatskyy, C. T. Lin, B. Keimer, and V. Hinkov, Phys. Rev. B \textbf{82}, 134503 (2010).

\bibitem{Diallo_PRB:2010} S. O. Diallo, D. K. Pratt, R. M. Fernandes, W. Tian, J. L. Zarestky, M. Lumsden, T. G. Perring, C. L. Broholm, N. Ni, S. L. Bud'ko, P. C. Canfield, H.-F. Li, D. Vaknin, A. Kreyssig, A. I. Goldman, and R. J. McQueeney, Phys. Rev.
B \textbf{81}, 214407 (2010).

\bibitem{Harriger_PRB:2011} L. W. Harriger, H. Q. Luo, M. S. Liu, C. Frost, J. P. Hu, M. R. Norman, and Pengcheng Dai, Phys. Rev.
B \textbf{84}, 054544 (2011).

\bibitem{Ewings_PRB:2011} R. A. Ewings, T. G. Perring, J. Gillett, S. D. Das, S. E. Sebastian, A. E. Taylor, T. Guidi, and A. T. Boothroyd, Phys. Rev. B \textbf{83}, 214519 (2011).

\bibitem{LiuNatPhys12} M. Liu, L. W. Harriger, H. Luo, M. Wang, R. Ewings, T. Guidi, H. Park, K. Haule, G. Kotliar, S. Hayden and P. Dai, Nat. Phys. \textbf{8}, 376 (2012).

\bibitem{Applegate} R. Applegate, J. Oitmaa, R. R. P. Singh, Phys. Rev. B {\bf 81}, 024505 (2010).


\bibitem{Lu_PRL:2018} X. Lu, D. D. Scherer, D. W. Tam, W. Zhang, R. Zhang, H. Luo, L. W. Harriger, H. C. Walker, D. T. Adroja, B. M. Andersen, and P. Dai, Phys. Rev. Lett. \textbf{121}, 067002 (2018).

\bibitem{Mila} A. L\"auchli, F. Mila, and K. Penc, Phys. Rev. Lett.
\textbf{97}, 087205 (2006).

\bibitem{footnote}
The parity is defined by the $\mathbb{Z}_{2}$ symmetry of the Hamiltonian
generated by $G\equiv\exp[i\pi\sum_{i}\tilde{S}_{i}^{z}]$.

\bibitem{Graser}
S. Graser, T. Maier, P. Hirschfeld, and D. Scalapino, New J. Phys. \textbf{11}, 025016 (2009).

\bibitem{Ran}
Y. Ran, F. Wang, H. Zhai, A. Vishwanath, and D.-H. Lee, Phys. Rev. B \textbf{79}, 014505 (2009).

\bibitem{Knolle}
J. Knolle, I. Eremin, A. Chubukov, and R. Moessner, Phys. Rev. B \textbf{81}, 140506(R) (2010).

\bibitem{Zhu10} J.-X. Zhu, R. Yu, H. Wang, L. L. Zhao, M. D. Jones, J. Dai, E. Abrahams, E. Morosan, M. Fang, and Q. Si, Phys. Rev. Lett. \textbf{104}, 216405 (2010).

\bibitem{Free10} D. G. Free and J. S. O. Evans, Phys. Rev. B \textbf{81}, 214433 (2010).

\bibitem{Freelon15} B. Freelon, Y. H. Liu, J.-L. Chen, L. Craco, M. S. Laad, S. Leoni, J. Chen, L. Tao, H. Wang, R. Flauca, Z. Yamani, M. Fang, C. Chang, J.-H. Guo, and Z. Hussain, Phys. Rev. B \textbf{92}, 155139 (2015).

\bibitem{MFang} M.-H. Fang, H.-D. Wang, C.-H. Dong, Z.-J. Li, C.-M. Feng, J. Chen, and H. Q. Yuan,
 Euro. Phys. Lett. \textbf{94}, 27009 (2011).

\bibitem{DMWang} D. M. Wang, J. B. He, T.-L. Xia and G. F. Chen,
Phys. Rev. B \textbf{83},132502 (2011).

\bibitem{Wang_PRB:2015} M. Wang, M. Yi, H. Cao, C. de la Cruz, S. K. Mo, Q. Z. Huang, E. Bourret-Courchesne, P. C. Dai, D. H. Lee, Z. X. Shen, and R. J. Birgeneau,
Phys. Rev. B \textbf{92}, 121101 (2015).

\bibitem{Song_NatComm:2016}
Y. Song, Z. Yamani, C. Cao, Y. Li, C. Zhang, J. S. Chen,
Q. Huang, H. Wu, J. Tao, Y. Zhu, W. Tian, S. Chi, H. Cao, Y.-B. Huang, M. Dantz, T. Schmitt, R. Yu, A. H. Nevidomskyy, E. Morosan, Q. Si and P. Dai, Nature Commun. {\bf 7}, 13879 (2016).

\bibitem{Song_PRL:2019}
Y. Song, H. Cao, B. C. Chakoumakos, Y. Zhao, A. Wang, H. Lei, C. Petrovic, and R. J. Birgeneau,
Phys. Rev. Lett. \textbf{122}, 087201 (2019).


\bibitem{Smerald2013}
A. Smerald and N. Shannon,
Phys. Rev. B \textbf{88}, 184430 (2013).



\bibitem{Barzykin}
V. Barzykin and L. P. Gor'kov,
Phys. Rev. Lett. \textbf{70}, 2479 (1993).



\bibitem{Portnichenko19}
P. Y. Portnichenko, S. E. Nikitin, A. Prokofiev, S. Paschen, J.-M. Mignot, J. Ollivier, A. Podlesnyak, Siqin Meng, Zhilun Lu,
and D. S. Inosov Phys. Rev. B \textbf{99}, 214431 (2019).

\bibitem{Paschen_2012} J. Custers, K. A. Lorenzer, M. M\"{u}ller, A. Prokofiev, A. Sidorenko, H. Winkler, A. M. Strydom, Y. Shimura, T. Sakakibara, Rong Yu, Q. Si, and S. Paschen, Nat. Mater. \textbf{11}, 189-194 (2012).
\end{thebibliography}

\end{document}